\documentclass{jpsj-suppl}
\usepackage{times}
\usepackage{color}

\usepackage{braket}
\usepackage{mathrsfs}

\usepackage{graphicx}
\usepackage{graphics}
\usepackage{amssymb, latexsym}
\usepackage{amsmath}
\usepackage{bm}
\usepackage{color}
\usepackage{ulem}
\usepackage{fancybox}
\usepackage{xspace}
\title{Linear Spin Wave Analysis for General Magnetic Orders
in the Kondo Lattice Model}

\author{Yutaka Akagi\thanks{E-mail address: akagi@aion.t.u-tokyo.ac.jp}, Masafumi Udagawa, and Yukitoshi Motome
}
\inst{Department of Applied Physics, University of Tokyo,
Hongo 7-3-1, Bunkyo-ku, Tokyo 113-8656 
}
\abst{
We extend the formulation of the spin wave theory for the Kondo lattice model, which was mainly used for the ferromagnetic metallic state, 
to general magnetic orders including complex noncollinear and noncoplanar orders. 
The $1/S$ expansion is reformulated in the matrix form depending on the size of the magnetic unit cell. 
The noncollinearity and noncoplanarity of the localized moments are properly taken into account by the matrix elements of the para-unitary matrix used in the diagonalization of the Bogoliubov-de Gennes type Hamiltonian for magnons.
We apply the formulation within the linear spin wave approximation to a typical noncollinear case, the $120^{\circ }$ N{\'e}el order on a triangular lattice at half filling. 
We calculate the magnon excitation spectrum and the quantum correction to the magnitude of ordered moments as functions of the strength of the Hund's-rule coupling, $J_{\rm H}/t$.
We find that the magnon excitation shows softening at $J_{\rm H}/t \simeq 2.9$, 
which indicates that the $120^{\circ }$ order is destabilized for smaller $J_{\rm H}/t$.
On the other hand, we show that the $120^{\circ }$ order is stable in the entire range of $J_{\rm H}/t \gtrsim 2.9$, and, in the limit of $J_{\rm H}/t \to \infty$,
the form of the spin wave spectrum approaches that for the antiferromagnetic Heisenberg model, while the bandwidth is proportional to $t^2/J_{\rm H}$. 
The reduction of the ordered moment is smaller than that for the spin-only model, except in the vicinity of the softening.
}

\kword{spin wave theory, Kondo lattice model, frustration, triangular lattice}

\begin{document}
\maketitle

\section{Introduction}
The Kondo lattice model is one of the fundamental models to describe the strongly correlated eletron systems
in which conduction electrons and localized spins interact with each other. 
For instance, the model with classical localized moments and the ferromagnetic spin-charge coupling (Hund's-rule coupling) was intensively studied for perovskite manganese oxides. 
The magneto-transport properties in experiments were well explained by the model, such as 
the transition to a ferromagnetic metallic state and the colossal magnetoresistance~\cite{Helmolt1993,Chahara1993,Jin1994,Furukawa1994,Furukawa1995,Kaplan1999,Tokura1999,Dagotto2001}. 
The spin excitation spectrum was also studied by the spin wave approximation, and the results were in good agreement with the neutron inelastic scattering experiments~\cite{Furukawa1996}.
In this class of spin-charge coupled systems, quantum spin fluctuations appear in a different form compared to those in localized spin systems, as the quantum effects are brought about through the coupling to conduction electrons. 
In fact, they lead to various nontrivial effects, such as a finite lifetime of magnon excitation even at $T=0$ appearing in higher-order corrections in the $1/S$ expansion ($S$ is the length of spin)~\cite{Golosov2000,Shannon2002}. 
An ``order from disorder" phenomenon was also found in the presence of the antiferromagnetic superexchange interaction between localized spins~\cite{Shannon2002}.

Recently, the Kondo lattice model has attracted renewed interest, since it gives rise to topologically nontrivial phases with noncoplanar magnetic orders~\cite{Ohgushi2000,Shindou2001,Martin2008,Akagi2010,Venderbos2012,Rahmani2013}. 
For instance, at $1/4$ and $3/4$ fillings of the Kondo lattice model on a triangular lattice, Chern insulating phases are stabilized with four-sublattice noncoplanar magnetic order~\cite{Martin2008,Akagi2010}.
The scalar chiral order in these phases affects the kinetic motion of itinerant electrons through the spin Berry phase, 
and gives rise to many fascinating transport phenomena, such as the topological Hall effect and the chiral edge current. 
The authors recently studied the effect of quantum spin fluctuations on the Chern insulating phases~\cite{Akagi2013_9}. 
The spin wave formulation for the ferromagnetic case was extended to the noncoplanar ordering. 
The extension is not limited to the four-sublattice case but widely applicable to general noncollinear and noncoplanar orders. 
The details of the extended framework, however, were not shown in the previous study. 
It is also desired to show other applications for demonstrating the extended method.

In this contribution, we provide the details of the spin wave formulation for general magnetic orders in the Kondo lattice model. 
In contrast to the simple ferromagnetic case, the magnon Green function is formulated in the matrix form, including the anomalous part.
Accordingly, the diagonalization of the magnon self-energy leads to the mixing of magnon creation and annihilation operators, resulting in the reduction of ordered magnetic moments.
We apply this extended method within the linear spin wave approximation to the $120^{\circ }$ N{\'e}el order on a triangular lattice at half filling.
As a result, we find that the magnon excitation spectrum exhibits softening at $J_{\rm H} \simeq 2.9$, which indicates an instability toward a different ordered state for smaller $J_{\rm H}$. 
On the other hand, in the limit of $J_{\rm H} \to \infty$, the spectrum form approaches that for the antiferromagnetic Heisenberg model with the effective exchange interaction $\sim t^2/J_{\rm H}$. 
We find that the reduction of the ordered moment is smaller than that for the Heisenberg model, except in the vicinity of the softening.

\section{Model and Method}
\label{sec:model}
We consider a simple Kondo lattice model in which conduction electrons interact with localized spins via the ferromagnetic Hund's-rule coupling. 
The Hamiltonian is given by 
\begin{eqnarray}
\mathcal{H} = -t\sum\limits_{\langle i\alpha, i'\alpha'\rangle}\sum\limits_{s=\uparrow, \downarrow}\bigl(c^{\dag}_{i\alpha s}c_{i'\alpha's} + {\rm H. c.}\bigr) 
- \frac{J_{\rm H}}{S}\sum\limits_{i,\alpha}\sum\limits_{s,s'=\uparrow, \downarrow}{\mathbf S}_{i\alpha}\cdot c^{\dag}_{i\alpha s}\bm{\sigma}_{ss'}c_{i\alpha s'},
\label{Ham}
\end{eqnarray}
where $c^{\dag}_{i\alpha s}$($c_{i\alpha s}$) is a creation (annihilation) operator for a conduction electron with spin $s$ at site $(i,\alpha)$. 
Here, we consider a general magnetic order in which the magnetic unit cell contains $n_{\rm sub}$ sites; $\alpha$ is the index for the sublattice ($\alpha=1$, $2$, $\cdots,$ $n_{\rm sub}$) and $i$ is the index for the unit cell. 
\textit{t} is the transfer integral and $J_{\rm H}(>0)$ is the Hund's-rule coupling (the sign of $J_{\rm H}$ does not change the results within the linear spin wave approximation),
$\bm{\sigma}_{ss'}=({\sigma}^x_{ss'},{\sigma}^y_{ss'},{\sigma}^z_{ss'})$
is a vector representation of Pauli matrices, and ${\mathbf S_{i\alpha}}$ 
denotes a localized spin at site $(i,\alpha)$ ($S$ is the magnitude of the spin).
The sum $\langle i\alpha, i'\alpha'\rangle$
is taken over the nearest-neighbor sites on the triangular lattice. 
Hereafter, we take $t=1$ as an energy unit, the lattice constant $a=1$, and the Planck constant divided by $2\pi$, $\hbar=1$.

In the spin wave calculations, we choose the spin quantization axis of itinerant electrons parallel to that for the ordered moments at each site. 
We define the ordered magnetic moments as 
$\langle{\mathbf S}_{i\alpha}\rangle = S (\sin\theta_{\alpha}\cos\phi_{\alpha}, \sin\theta_{\alpha}\sin\phi_{\alpha}, \cos\theta_{\alpha})$ at each sublattice. 
We denote the electron spin state parallel (anti-parallel) to the localized moment as $\ket{+_{\alpha}} (\ket{-_{\alpha}})$:
\begin{eqnarray}
\left\{\begin{array}{ll}
\displaystyle \ket{+_{\alpha}} = \cos\frac{\theta_{\alpha}}{2}\ket{\uparrow} + e^{i\phi_{\alpha}}\sin\frac{\theta_{\alpha}}{2}\ket{\downarrow}\\
\displaystyle \ket{-_{\alpha}} =  -e^{-i\phi_{\alpha}}\sin\frac{\theta_{\alpha}}{2}\ket{\uparrow} + \cos\frac{\theta_{\alpha}}{2}\ket{\downarrow},
\label{eq:rotated_frame}
\end{array}\right.
\end{eqnarray}
where $\ket{\uparrow} = c_{i\alpha \uparrow}^\dagger |0\rangle$ and 
$\ket{\downarrow} = c_{i\alpha \downarrow}^\dagger |0\rangle$ 
[$|0\rangle$ is the empty state at $(i,\alpha)$]. 
In this local frame, the Hamiltonian in Eq.~(\ref{Ham}) is written as
\begin{eqnarray}
\mathcal{H} = -t\!\!\!\sum\limits_{\langle i\alpha, i'\alpha'\rangle}\!\sum\limits_{s,s'=\pm}\!\!\bigl( \braket{s_{\alpha}|s'_{\alpha'}} \tilde{c}^{\dag}_{i\alpha s}\tilde{c}_{i'\alpha' s'} + {\rm H. c.}\bigr) 
- \frac{J_{\rm H}}{S} \sum\limits_{i,\alpha}\sum\limits_{s,s'=\pm}\tilde{\mathbf S}_{i\alpha}\cdot \tilde{c}^{\dag}_{i\alpha s}\bm{\sigma}_{ss'}\tilde{c}_{i\alpha s'},
\label{Ham2}
\end{eqnarray}
where
$\tilde{c}_{i\alpha s}$($\tilde{c}_{i\alpha s}^\dagger$)
is the operator to annihilate (create) an electron in the states defined in Eq.~(\ref{eq:rotated_frame}), and $\langle\tilde{\mathbf S}_{i\alpha}\rangle = (0,0,S)$.

In order to consider the effect of quantum fluctuations of localized moments in the spin wave approximation, 
we apply the Holstein-Primakoff transformation in the lowest order of $1/S$,
\begin{eqnarray}
\tilde{S}^+_{i\alpha} \simeq \sqrt{2S}a_{i\alpha}, \quad 
\tilde{S}^-_{i\alpha} \simeq \sqrt{2S}a^{\dag}_{i\alpha}, \quad 
\tilde{S}^z_{i\alpha} = S - a^{\dag}_{i\alpha}a_{i\alpha},
\label{Holstein-Primakoff}
\end{eqnarray}
to the Hamiltonian in Eq.~(\ref{Ham2}). 
Here, $a^{\dag}_{i\alpha}$($a_{i\alpha}$) is the magnon creation (annihilation) operator at site $(i,\alpha)$.
Then, we divide the transformed Hamiltonian into two parts, 
$
\mathcal{H} = \mathcal{H}_0 + \mathcal{H}' 
$:
\begin{align}
\begin{array}{ll}
\displaystyle \mathcal{H}_0\! =\! -t\!\!\!\sum\limits_{\langle i\alpha, i'\alpha'\rangle}\!\sum\limits_{s,s'=\pm}\!\!\bigl( \braket{s_{\alpha}|s'_{\alpha'}} \tilde{c}^{\dag}_{i\alpha s}\tilde{c}_{i'\alpha' s'}\! + {\rm H. c.}\bigr) \
- J_{\rm H}\sum\limits_{i,\alpha}(\tilde{c}^{\dag}_{i\alpha+}\tilde{c}_{i\alpha+}\! -\! \tilde{c}^{\dag}_{i\alpha-}\tilde{c}_{i\alpha-})\\
\displaystyle \mathcal{H}'\!\! =\! -J_{\rm H}\!\sum\limits_{i,\alpha}\Bigl[\sqrt{\frac{2}{S}}(a_{i\alpha}\tilde{c}^{\dag}_{i\alpha-}\tilde{c}_{i\alpha+}\!\! +\! a^{\dag}_{i\alpha}\tilde{c}^{\dag}_{i\alpha+}\tilde{c}_{i\alpha-}) 
- \frac{1}{S} a^{\dag}_{i\alpha}a_{i\alpha}(\tilde{c}^{\dag}_{i\alpha+}\tilde{c}_{i\alpha+}\!\! -\! \tilde{c}^{\dag}_{i\alpha-}\tilde{c}_{i\alpha-})\Bigr].
\label{perturbation}
\end{array} 
\end{align}
Here, 
$\mathcal{H}_0$ describes the interaction between itinerant electrons and static ordered moments, and 
$\mathcal{H}'$ denotes the interaction between electrons and magnons, which is composed of the higher order terms in $1/S$.
Note that $\mathcal{H}_0$ corresponds to the saddle point Hamiltonian in the variational study~\cite{Akagi2010},
and we consider the quantum corrections from $\mathcal{H}'$ systematically below.

We here perform a perturbation expansion in terms of $\mathcal{H}'$.
By generalizing the ferromagnetic case~\cite{Furukawa1996},
we introduce the magnon Green function in the matrix form; 
\begin{eqnarray}
\hat{\bm D}_{{\mathbf q}}
(\tau) = \begin{bmatrix}
D^{++}_{{\mathbf q}\alpha_1 \alpha_2}(\tau) &\!\!\!\!D^{+-}_{{\mathbf q}\alpha_1 \alpha_2}(\tau)\\
D^{-+}_{{\mathbf q}\alpha_1 \alpha_2}(\tau) &\!\!\!\!D^{--}_{{\mathbf q}\alpha_1 \alpha_2}(\tau)
\end{bmatrix} 
=\begin{bmatrix}
-\langle T_\tau a_{{\mathbf q}\alpha_1}(\tau)a^{\dag}_{{\mathbf q}\alpha_2}(0)\rangle &\!\!\!\!\!-\langle T_\tau a_{{\mathbf q}\alpha_1}(\tau)a_{-{\mathbf q}\alpha_2}(0)\rangle\\
-\langle T_\tau a^{\dag}_{-{\mathbf q}\alpha_1}(\tau)a^{\dag}_{{\mathbf q}\alpha_2}(0)\rangle &\!\!\!\!\!-\langle T_\tau a^{\dag}_{-{\mathbf q}\alpha_1}(\tau)a_{-{\mathbf q}\alpha_2}(0)\rangle
\end{bmatrix},
\end{eqnarray}
where each $D^{\pm\pm}$ is the $n_{\rm sub}\times n_{\rm sub}$ matrix in terms of the sublattice indices $\alpha_1$ and $\alpha_2$ ($D^{+-}$ and $D^{-+}$ are anomalous Green functions),
$T_\tau$ represents time-ordered product,
${\mathbf q}$ is a wave vector, and $\tau$ is an imaginary time.
The matrix form of the Dyson equation for $\hat{\bm D}_{{\mathbf q}}$ is given by 
\begin{eqnarray}
\hat{{\bm D}}^{-1}_{\mathbf q}(i\omega_{n}) = \hat{{\bm D}}^{(0)-1}_{\mathbf q}(i\omega_{n}) - \hat{{\bm \Sigma}}_{\mathbf q}(i\omega_{n}),
\label{Dyson_eq}
\end{eqnarray}
where $\hat{{\bm D}}^{(0)}_{\mathbf q}(i\omega_{n}) = (1/i\omega_{n}) \hat{{\bm \tau}}$ 
is the bare magnon Green function, 
$\hat{{\bm \Sigma}}_{\mathbf q}(i\omega_n)$ is the magnon self-energy,
$\omega_n = 2\pi n/\beta$ is the bosonic Matsubara frequency ($n$ is an integer and $\beta$ is inverse temperature),
and 
$\hat{{\bm \tau}} = 
\begin{bmatrix} \hat{{\mathbf 1}} & \hat{{\mathbf 0}} \\
\hat{{\mathbf 0}} & -\hat{{\mathbf 1}} \end{bmatrix}$ ($\hat{{\mathbf 1}}$ and $\hat{{\mathbf 0}}$ are the $n_{\rm sub}\times n_{\rm sub}$ unit matrix and null matrix, respectively).

We expand the self-energy $\hat{{\bm \Sigma}}_{\mathbf q}(i\omega_n)$ in terms of $1/S$, in order to take account of the quantum fluctuation systematically. Up to the order of $1/S$, the matrix elements of the self-energy are given as follows:
\begin{align}
&\Sigma^{++}_{{\mathbf q}\alpha_1\alpha_2}(i\omega_n) = \delta_{\alpha_1\alpha_2}\frac{J_{\rm H}}{S N_{\rm unit}}\sum_{\mathbf k}(n_{{\mathbf k}\alpha_1+} - n_{{\mathbf k}\alpha_1-})\nonumber\\ 
& \qquad +\frac{2J_{\rm H}^2}{S N_{\rm unit}}\sum_{\mathbf k}e^{i{\mathbf G}\cdot({\mathbf e}_{\alpha_1} - {\mathbf e}_{\alpha_2})}\sum_{\eta,\eta'}\frac{f_{\rm F}(\epsilon_{{\mathbf k},\eta'}-\mu) -f_{\rm F}(\epsilon_{{\mathbf k}+{\mathbf q}+{\mathbf G},\eta}-\mu)}{\epsilon_{{\mathbf k},\eta'} -\epsilon_{{\mathbf k}+{\mathbf q}+{\mathbf G},\eta} +i\omega_n}\nonumber\\
& \qquad \times\braket{-_{\alpha_1}|{\mathbf k}+{\mathbf q}+{\mathbf G},\eta} \braket{{\mathbf k}+{\mathbf q}+{\mathbf G},\eta|-_{\alpha_2}} 
\braket{+_{\alpha_2}|{\mathbf k},\eta'} \braket{{\mathbf k},\eta'|+_{\alpha_1}}\Bigg|_{{\mathbf k}+{\mathbf q}+{\mathbf G}\in{\rm 1st. BZ}},
\end{align}
\begin{align}
&\Sigma^{+-}_{{\mathbf q}\alpha_1\alpha_2}(i\omega_n) = \frac{2J_{\rm H}^2}{S N_{\rm unit}}\sum_{\mathbf k}e^{i{\mathbf G}\cdot({\mathbf e}_{\alpha_1} - {\mathbf e}_{\alpha_2})}\sum_{\eta,\eta'}\frac{f_{\rm F}(\epsilon_{{\mathbf k},\eta'} -\mu) -f_{\rm F}(\epsilon_{{\mathbf k}+{\mathbf q}+{\mathbf G},\eta} -\mu)}{\epsilon_{{\mathbf k},\eta'} - \epsilon_{{\mathbf k}+{\mathbf q}+{\mathbf G},\eta} +i\omega_n}\nonumber\\
& \qquad \times\braket{-_{\alpha_1}|{\mathbf k}+{\mathbf q}+{\mathbf G},\eta} \braket{{\mathbf k}+{\mathbf q}+{\mathbf G},\eta|+_{\alpha_2}} 
\braket{-_{\alpha_2}|{\mathbf k},\eta'} \braket{{\mathbf k},\eta'|+_{\alpha_1}}\Bigg|_{{\mathbf k}+{\mathbf q}+{\mathbf G}\in{\rm 1st. BZ}},
\end{align}
where $\mu$ is the chemical potential,
$f_{\rm F}$ is the Fermi distribution function,
$N_{\rm unit}$ is the number of unit cells in the system, 
${\mathbf e}_{\alpha}$ corresponds to the internal position of the $\alpha$-sublattice site within the magnetic unit cell; 
the sum over ${\mathbf k}$ is taken so that ${\mathbf k}+{\mathbf q}+{\mathbf G}$ is in the first Brillouin zone with a reciprocal vector ${\mathbf G}$. 
Here, $\braket{s_{\alpha}|{\mathbf k},\eta}$ is the Bloch wave function which satisfies the eigenvalue equation 
\begin{equation}
\sum_{\alpha'=1}^{n_{\rm sub}}\sum_{s'=\pm}
\bra{s_{\alpha}}{\cal H}_0({\mathbf k}) \ket{s'_{\alpha'}}\braket{s'_{\alpha'}|{\mathbf k},\eta} = \epsilon _{{\mathbf k},\eta }\braket{s_{\alpha}|{\mathbf k},\eta} 
\end{equation}
with the unperturbed Hamiltonian ${\cal H}_0({\mathbf k})$ in the momentum representation, and
$n_{{\mathbf k}\alpha s} = \sum_{\eta}|\braket{s_{\alpha}|{\mathbf k}, \eta}|^2f_{\rm F}(\epsilon_{{\mathbf k},\eta}-\mu)$ 
is the occupation number of itinerant electrons of the state characterized by the indices ${\mathbf k}$, $\alpha$, and $s$.
The other matrix elements of the self-energy are obtained by the relations, 
$\Sigma^{--}_{{\mathbf q}\alpha_1 \alpha_2}(i\omega_n) = \Sigma^{++*}_{-{\mathbf q}\alpha_1 \alpha_2}(i\omega_n)$ and
$\Sigma^{-+}_{{\mathbf q}\alpha_1 \alpha_2}(i\omega_n) = \Sigma^{+-*}_{-{\mathbf q}\alpha_1 \alpha_2}(i\omega_n)$. 
Furthermore, the matrix elements satisfy the relations,
$\Sigma^{++}_{{\mathbf q}\alpha_1 \alpha_2}(i\omega_n) = \Sigma^{++*}_{{\mathbf q}\alpha_2 \alpha_1}(-i\omega_n)$,
$\Sigma^{+-}_{{\mathbf q}\alpha_1 \alpha_2}(i\omega_n) = \Sigma^{-+*}_{{\mathbf q}\alpha_2 \alpha_1}(-i\omega_n)$, and
$\Sigma^{--}_{{\mathbf q}\alpha_1 \alpha_2}(i\omega_n) = \Sigma^{--*}_{{\mathbf q}\alpha_2 \alpha_1}(-i\omega_n)$; hence,
$\hat{{\bm \Sigma}}_{\mathbf q}(0)$ is a Hermitian matrix.

The magnon spectrum is obtained by making analytic continuation, 
$i\omega_n\rightarrow\omega + i\delta$, 
and tracing the poles of Green function, i.e., the zeros of $\hat{{\bm D}}^{-1}_{\mathbf q}(\omega) = \omega \hat{{\bm \tau}} - \hat{{\bm \Sigma}}_{\mathbf q}(\omega)$. 
At the lowest order of $1/S$, the self-energy can be approximated by its static value at $\omega=0$, and the magnon spectrum is obtained from the eigenvalues of 
$\hat{{\bm \tau}}\hat{{\bm \Sigma}}_{\mathbf q}(0)$~\cite{Furukawa1996}. 
Thus, the eigenvalue equation is written as
\begin{eqnarray}
\hat{{\bm T}}^{-1}_{\mathbf q} \hat{{\bm \tau}} \hat{{\bm \Sigma}}_{\mathbf q}(0) \hat{{\bm T}}_{\mathbf q} =\hat{{\bm \tau}} \hat{{\bm \lambda}}_{\mathbf q},
\label{diagonalization}
\end{eqnarray}
where $\hat{{\bm \lambda}}_{\mathbf q}$ is the eigenvalue (matrix) of the magnon self-energy,
and $\hat{{\bm T}}_{\mathbf q}$ is the para-unitary matrix which satisfies the condition 
$\hat{{\bm T}}^{\dag}_{\mathbf q}\hat{{\bm \tau}} \hat{{\bm T}}_{\mathbf q} = \hat{{\bm \tau}}$~\cite{Colpa1978}.
Note that the eigenvalue problem has the same structure as that for the Bogoliubov-de Gennes type Hamiltonian in bosonic systems.

The effect of zero-point oscillation to the length of ordered moment is also obtained by the spin wave formulation introduced above. 
From Eq.~(\ref{Holstein-Primakoff}), the reduction of ordered moment is given by
$\Delta S_{\alpha}=S-\langle \tilde{S}^z_{i\alpha}\rangle =\langle a^{\dag}_{i\alpha}a_{i\alpha}\rangle
$,
which is calculated as
\begin{eqnarray}
\Delta S_{\alpha}\!\!\!\!\!\! &=&\!\!\!\!\!\!
\frac{1}{N_{\rm unit}}\sum_{i} \langle a^{\dag}_{i\alpha} a_{i\alpha} \rangle 
=\frac{-1}{N_{\rm unit}}\sum\limits_{{\mathbf q}} \hat{D}^{++}_{{\mathbf q}\alpha\alpha}(-\delta) \notag\\
&=&\!\!\!\!\!\!\frac{-1}{N_{\rm unit}}\sum\limits_{{\mathbf q}} \frac{1}{\beta} \sum\limits_{i\omega_n} e^{i\omega_n \delta} \hat{D}^{++}_{{\mathbf q}\alpha\alpha}(i\omega_n) 
=\frac{1}{N_{\rm unit}}\sum\limits_{{\mathbf q}} \sum_{\eta=1}^{n_{\rm sub}}
\frac{|T_{{\mathbf q}\alpha\eta}|^2}{e^{\beta \lambda_{{\mathbf q}\eta}}-1} {\rm sign}({\lambda_{{\mathbf q}\eta}}),
\label{moment-reduction}
\end{eqnarray}
where $\delta (>0)$ is an infinitesimal,
${\rm sign}(x)=1 (-1)$ for $x>0$ ($x<0$). 
In the last line of Eq.~(\ref{moment-reduction}), we used the following equation,
\begin{eqnarray}
\hat{{\bm D}}_{\mathbf q}(i\omega_n) &=& \hat{{\bm T}}_{\mathbf q} [i\omega_n \hat{{\mathbf 1}} - \hat{{\bm \tau}} \hat{{\bm \lambda}}_{\mathbf q} ]^{-1} \hat{{\bm T}}_{\mathbf q}^{-1} \hat{{\bm \tau}},
\end{eqnarray}
which is obtained from the Dyson equation in Eq.~(\ref{Dyson_eq}) by using Eq.~(\ref{diagonalization}).

\section{Results}
\begin{figure}[t]
\begin{center}
\includegraphics[width=16.0cm]{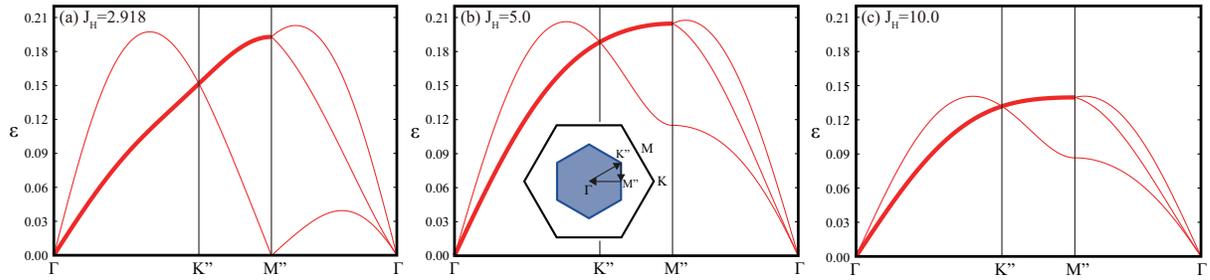}
\end{center}
\caption{(Color online) Magnon excitation spectra in the $120^{\circ }$ N{\'e}el ordered state at half filling for the Kondo lattice model in Eq.~(\ref{Ham}) at (a) $J_{\rm H}=2.918$, (b) $5.0$, and (c) $10.0$.
The thin (thick) curves denote the non(doubly)-degenerate branches.
The blue (white) hexagon in the inset of (b) indicates the folded (original) Brillouin zone.
The magnon dispersions are plotted along the symmetric lines in the folded Brillouin zone.
}
\label{Fig_dispersion}
\end{figure}

As a demonstration of the extended formulation of the spin wave theory, 
we here apply it to a typical noncollinear order, the three-sublattice $120^{\circ }$ N{\'e}el order on a triangular lattice.
In the variational calculation for the ground state of the Kondo lattice model in Eq.~(\ref{Ham}),
the $120^{\circ }$ N{\'e}el order is widely seen for all $J_{\rm H}$ at half filling (see Ref.~\cite{Akagi2010}). 
Setting 
$\langle{\mathbf S}_{i\alpha}\rangle = S (\sin\theta_{\alpha}\cos\phi_{\alpha}, \sin\theta_{\alpha}\sin\phi_{\alpha}, \cos\theta_{\alpha})$
with $\theta_1=\theta_2=\theta_3=\pi/2$, $\phi_1=0$, $\phi_2=2\pi/3$, and $\phi_3=-2\pi/3$,
and the internal coordinates as ${\mathbf e}_1 = (0, 0)$, ${\mathbf e}_2 = (1, 0)$, and ${\mathbf e}_3 = (1/2, \sqrt{3}/2)$,
we calculate the magnon dispersion in the $120^{\circ }$ N{\'e}el ordered state in the Kondo lattice model at half filling by the procedure introduced in Sec.~\ref{sec:model}. 

Figure~\ref{Fig_dispersion} shows the results of the magnon excitation spectra. 
The magnon excitation has three gapless modes, 
which are Nambu-Goldstone modes coming from the breaking of $SO(3)$ symmetry. 
As decreasing $J_{\rm H}$, the excitation spectrum changes gradually and shows softening at (or very nearby) the M" point for $J_{\rm H} \simeq 2.9$, 
as shown in Fig.~\ref{Fig_dispersion}(a).
This is a signature of the instability toward a different ordered state with a larger magnetic unit cell (possibly a 12-sublattice order).
On the other hand, the magnon excitation does not show any softening for larger $J_{\rm H}$. 
As increasing $J_{\rm H}$, the form of the magnon excitation spectrum approaches that for the Heisenberg model with the effective antiferromagnetic nearest-neighbor exchange interactions $\sim t^2/J_{\rm H}$ [see Fig.~\ref{Fig_dispersion}(c)]. 
%Similar asymptotic behavior was also found in the ferromagnetic case~\cite{Furukawa2003}. 
We note that, in the ferromagnetic case, similar asymptotic behavior was found, while the effective interaction is ferromagnetic and proportional to $t$~\cite{Furukawa2003}.

\begin{figure}[h]
\begin{center}
\includegraphics[width=7.0cm]{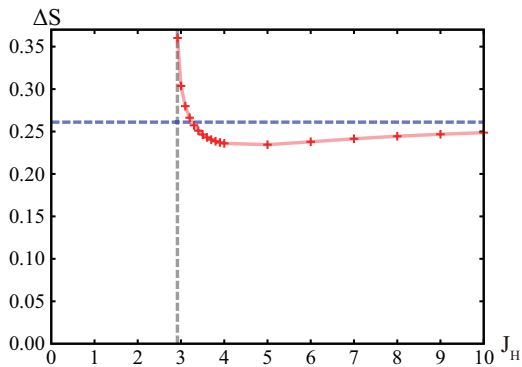}\hspace{2pc}
\begin{minipage}[b]{14pc}\caption{
\label{Fig_moment-reduction}
(Color online) The moment reduction of the $120^{\circ }$ N{\'e}el order at zero temperature. 
The vertical dashed line denotes the values of $J_{\rm H}$ at which the magnon excitation shows the softening.
The horizontal dashed line indicates the value of moment reduction of the $120^{\circ }$ N{\'e}el order in the antiferromagnetic Heisenberg model.
}
\end{minipage}
\end{center}
\end{figure}

We also calculate the effect of zero-point oscillation on the length of ordered moment in the $120^{\circ }$ N{\'e}el ordered state.
In the current case, the ordered moments are uniformly reduced at all sublattices by $\Delta S \equiv \Delta S_\alpha$. 
Figure~\ref{Fig_moment-reduction} shows the result for $\Delta S$ calculated by Eq.~(\ref{moment-reduction}).
As shown in the figure, $\Delta S$ in the large $J_{\rm H}$ limit approaches the value of moment reduction in the Heisenberg model, $\Delta S \simeq 0.261$~\cite{Jolicoeur1999}. 
This is consistent with the asymptotic behavior of the magnon spectrum observed in Fig.~\ref{Fig_dispersion}. 
As decreasing $J_{\rm H}$, $\Delta S$ decreases monotonically, except for the enhancement near $J_{\rm H} \simeq 2.9$ as a precursor of the softening. 
($\Delta S$ does not diverge even right at the softening in the two-dimensional case.)
The result indicates that the moment reduction by the quantum fluctuation through the spin-charge coupling becomes smaller compared to that in the spin-only model. 
A similar tendency is also seen for other orders, such as the collinear N{\'e}el order on a square lattice at half filling. 

\section{Summary}
To summarize, we have provided the detailed formulation of the spin wave theory for general magnetic orders in the Kondo lattice model.
We have applied this formulation within the linear spin wave approximation to the $120^{\circ }$ N{\'e}el order on a triangular lattice at half filling. 
As a result, we have found that the order is destabilized by the quantum fluctuation at $J_{\rm H} \simeq 2.9$.
We have also shown that, in the large $J_{\rm H}$ limit, the magnon dispersion and the moment reduction approach those in the antiferromagnetic Heisenberg model 
with the effective exchange interaction $\sim t^2/J_{\rm H}$. 
Our results indicate that the reduction of the ordered moment becomes smaller than that for the Heisenberg model, except in the region close to the softening.

Although we performed the spin wave analysis in the lowest order of $1/S$ expansions, higher-order quantum corrections may lead to qualitatively new effects~\cite{Golosov2000,Shannon2002}. 
We are also interested in quantum fluctuation effects on the competition between the double-exchange ferromagnetism and other elements, 
such as the antiferromagnetic superexchange interaction, single ion anisotropy, and external magnetic field,
which gives rise to some nontrivial magnetic orders even in the case of classical localized spins~\cite{Akagi2011,Akagi2013_preparation}.
These are left for future study.
\\

\noindent
{\large \textbf{Acknowledgement}}
\\

The authors acknowledge helpful discussions with Cristian D. Batista, Takahiro Misawa, Joji Nasu, Nic Shannon, and Youhei Yamaji.
Y.A. is supported by Grant-in-Aid for JSPS Fellows.
This work was supported by Grants-in-Aid for Scientific Research (Grants No. 24340076 and 24740221),
the Strategic Programs for Innovative Research (SPIRE),
MEXT, and the Computational Materials Science Initiative (CMSI), Japan.

\end{document}